\documentclass[conference]{IEEEtran}
\IEEEoverridecommandlockouts

\usepackage{cite}
\usepackage{amsmath,amssymb,amsfonts}
\usepackage{algorithmic}
\usepackage{textcomp}
\usepackage{svg}
\usepackage{amsmath}
\usepackage{xcolor}
\usepackage{graphicx}
\usepackage{physics}
\usepackage{graphics}
\usepackage{float}
\usepackage{cclicenses}
\usepackage{xspace}
\usepackage{subfigure}
\usepackage{makecell}
\usepackage{multirow} 
\usepackage[T1]{fontenc}

\usepackage[ruled , linesnumbered]{algorithm2e}
\usepackage{cleveref}

\begin{document}
\title{Quantum Key Distribution with Minimal Qubit Transmission Based on Multi-Qubit Greenberger–Horne–Zeilinger State}
\author{\IEEEauthorblockN{Tasdiqul Islam and Engin Arslan}
\IEEEauthorblockA{University of Nevada Reno\\
tasdiqul@nevada.unr.edu, earslan@unr.edu}}

\maketitle

\newcommand{\name}{$QKD$\xspace}

\begin{abstract}
Conventional Quantum Key Distribution (QKD) requires the transmission of multiple qubits equivalent to the length of the key. As quantum networks are still in their infancy thus, they are expected to have a limited capacity, necessitating too many qubit transmissions for QKD might limit the effective use of limited network bandwidth of quantum networks. To address this challenge and enhance the practicality of QKD, we propose a Multi-Qubit Greenberger–Horne–Zeilinger (GHZ) State-based QKD scheme that requires a small number of qubit transmissions. The proposed method transmits one qubit between endpoints and reuses it for the transmission of multiple classical bits with the help of Quantum nondemolition (QND) measurements. We show that one can transfer L-1 classical bits by generating an L-qubit GHZ state and transferring one to the remote party. We further show that the proposed QKD algorithm can be extended to enable multi-party QKD. It can also support QKD between parties with minimal quantum resources. As a result, the proposed scheme offers a quantum network-efficient alternative QKD.
\end{abstract}

\maketitle

\section{Introduction}
Quantum Key Distribution (QKD) is one of the practical use cases for quantum communication that offers a highly secure exchange of encryption keys between end users~\cite{bennett2020quantumbb84,ekert1991quantumE91,bennett1992quantumB92}. The most fundamental QKD algorithms such as BB84~\cite{bennett2020quantumbb84}, B92~\cite{bennett1992quantumB92}, and E91~\cite{ekert1991quantumE91} require the number of transferred qubits to be equal or larger than the size of the secret key. This, in turn, requires high-capacity quantum channels and quantum repeaters to accommodate qubit transmissions of all users in the network. However, the advancement of qubit transmission and repeater designs is still in its early stages, implying that the development of high-capacity quantum networks will likely require time. 

To overcome potential bandwidth limitations of quantum networks while still offering the critical QKD service, we introduce a QKD scheme with minimum qubit transmission with the help of a multi-qubit Greenberger-Horne-Zeilinger (GHZ) state. In the proposed algorithm, if Alice wants to share a key in $L$ length with Bob, she generates $L +1$ GHZ state entangled qubits and sends one of them to Bob. Alice then encodes an ancillary bit based on the value of the first bit in the key and teleports it to Bob by conducting Bell State Measurements (BSM) with the ancillary bit and the first qubit of remaining $L$ bit GHZ state entangled qubits in Alice. Upon receiving the BSM results, Bob performs the corresponding gate operations before conducting Quantum nondemolition (QND) measurement. The QND returns the probability distribution of the ancillary qubit which is used to infer the value of the first bit of the key. Next, Alice and Bob execute a series of gates to reverse the impact of the BSM conducted to transmit the first classical bit. Finally, they repeat the process for the next bit in the key until the key is fully transmitted. 

The encoding of the ancillary bit is important to ensure that the impact of BSM can be reversed which is critical to reset the values of qubits in the GHZ state. Specifically, if the value of the classical bit in the key is $0$, then Alice chooses $\alpha>\beta$ when encoding the ancillary bit in state $\alpha_1\ket{0} + \beta\ket{1}$. If the classical bit is $1$, then the values of $\alpha$ and $\beta$ are chosen to ensure that $\alpha<\beta$. Such encoding helps Bob to learn the value of the classical bit in the key without measuring its qubit. The proposed model can transmit $L$ length key by transmitting only one qubit (or as small as possible transmissions in the case of channel noise), significantly minimizing the number of qubit transmissions thereby reducing the load on quantum networks. It is worth noting that this scheme necessitates multi-qubit entanglement on Alice's side and classical communication to transfer the BSM results. 
In addition to offering a strong key exchange mechanism for two-party, we also show that the proposed solution can be extended to multiparty quantum key distribution. It also can be extended as server-client architecture where two clients with small capacity can share large keys with the help of a trusted server with high qubit generation capacity. Our contributions can be summarized as follows:

\begin{itemize}
    \item We propose a new Quantum Key Distribution scheme that requires only one qubit transmission between end users. The proposed method relies on Multi-Qubit Greenberger–Horne–Zeilinger (GHZ) State to create entangled qubits and specially encoded ancillary qubits to transfer the classical bits of a given key. 
    \item We demonstrate that the proposed multi-qubit GHZ can significantly reduce the need for qubit transmissions in QKD compared to the state-of-the-art methods. 
    \item We describe potential attacks and discuss how they can be prevented using available methods such as CHSH inequlities.
\end{itemize}

\section{Related Work}\label{sec:related}
QKD is one of the most popular applications of quantum networking. In 1984, Bennett and Brassard proposed a QKD scheme known as BB84  based on a single polarized photon~\cite{bennett2020quantumbb84}. Then, in 1991, Ekert experimented with QKD based on Bell's theorem \cite{ekert1991quantumE91}. Entanglement-based QKD was introduced in 1992 by Bennett \cite{bennett1992quantumB92}. There are many implementations and improvements for QKD. In~\cite{koashi1997quantum,cabello2000quantum}, authors tried to improve the efficiency of QKD so that there is less qubit transmission in exchange for more classical bit transmission. Cabello \cite{cabello2000quantum} attempted to enhance the protocol by improving the measurement strategy, while Koashi \cite{koashi1997quantum} enhanced the security and performance by updating the bit communication technique. Our work shares a similar objective with theirs; however, we employ teleportation and Quantum Non-Demolition (QND) as our primary components. Quantum secret sharing between multi-party is also an important branch of QKD. Li et al. used GHZ state to minimize the quantum resources for multi-party quantum secret sharing \cite{li2022multipartyGHZ}. Qin et al. proposed  multi-party quantum secret sharing also using GHZ state where multiple participants can be added or removed \cite{qin2017dynamicGHZ}. Cardoso-Isidoro proposed a scheme for QKD based on asymmetric double quantum teleportation \cite{cardoso2022sharedQKD_Tel}. In \cite{pryde2004measuringQND}, authors proposed a scheme for QND measurement for photonic qubits. In \cite{ralph2004quantumQNDper}, Ralph discussed the characterization and properties of different QND measurements. Similarly, we rely on Bob having a quantum memory with long enough storage time to execute QND measurements to the same qubit for consecutive bit transmissions. Ma et al. showed that quantum memories could store qubits for up to an hour~\cite{ma2021one1hour}; thus we believe our approach is feasible. 

In \cite{zhao2021GHZcreation2000}, Zhao et al. proposed an entanglement-creation scheme that can create $2,000$-atom GHZ states with more than $80\%$ fidelity.  In \cite{mooney2021generation27}, Mooney et al. described the creation of a 27 qubit GHZ state,  and in \cite{mooney2021whole65}, they described the creation of a GHZ state with $65$ qubits. As a result, we believe that our proposed mechanism that relies on multi-qubit GHZ states is feasible and can significantly reduce the need for qubit transmissions for QKD. 

\section{The System Model}
\begin{figure*}
\begin{center}
\subfigure[Step 1: GHZ State Preparation]{
\frame{\includegraphics[keepaspectratio=true,angle=0,width=.30\linewidth] {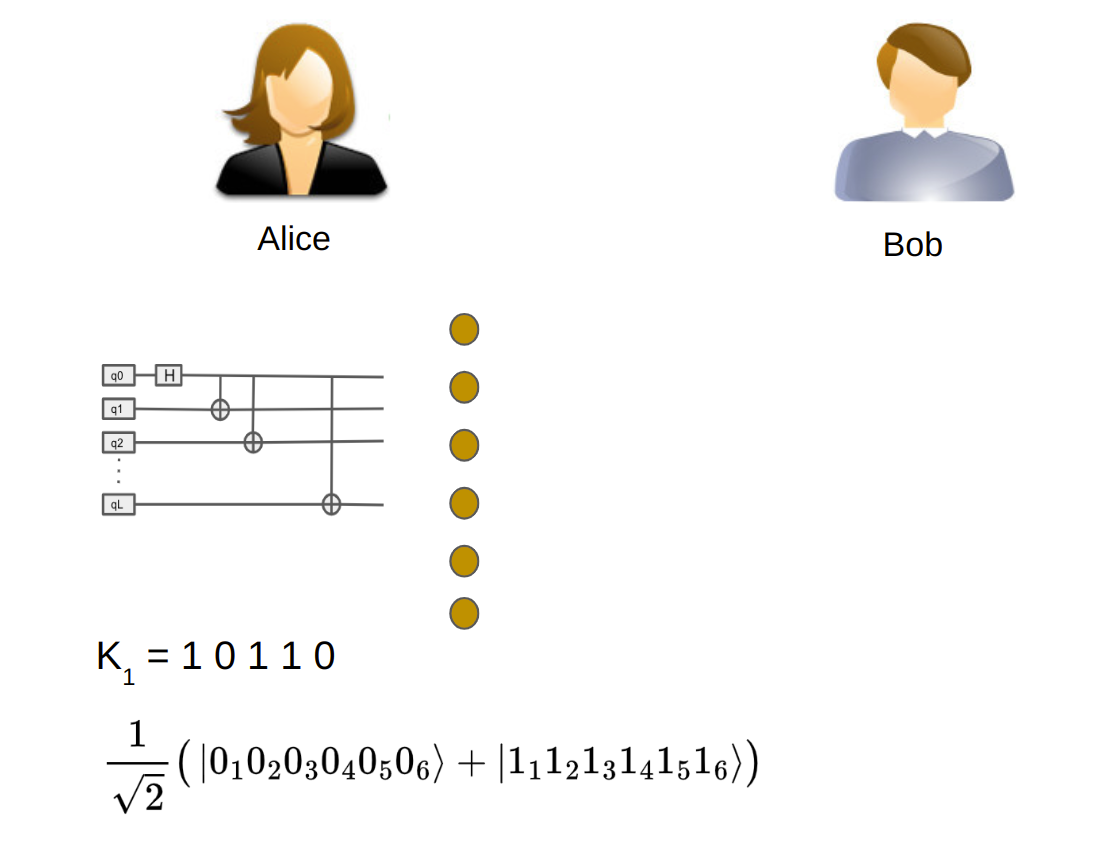}}
\label{fig:step1}}
\hspace{-4mm}
\subfigure[Step 2:Qubit Transmission to Bob]{
\frame{\includegraphics[keepaspectratio=true,angle=0,width=.30\linewidth] {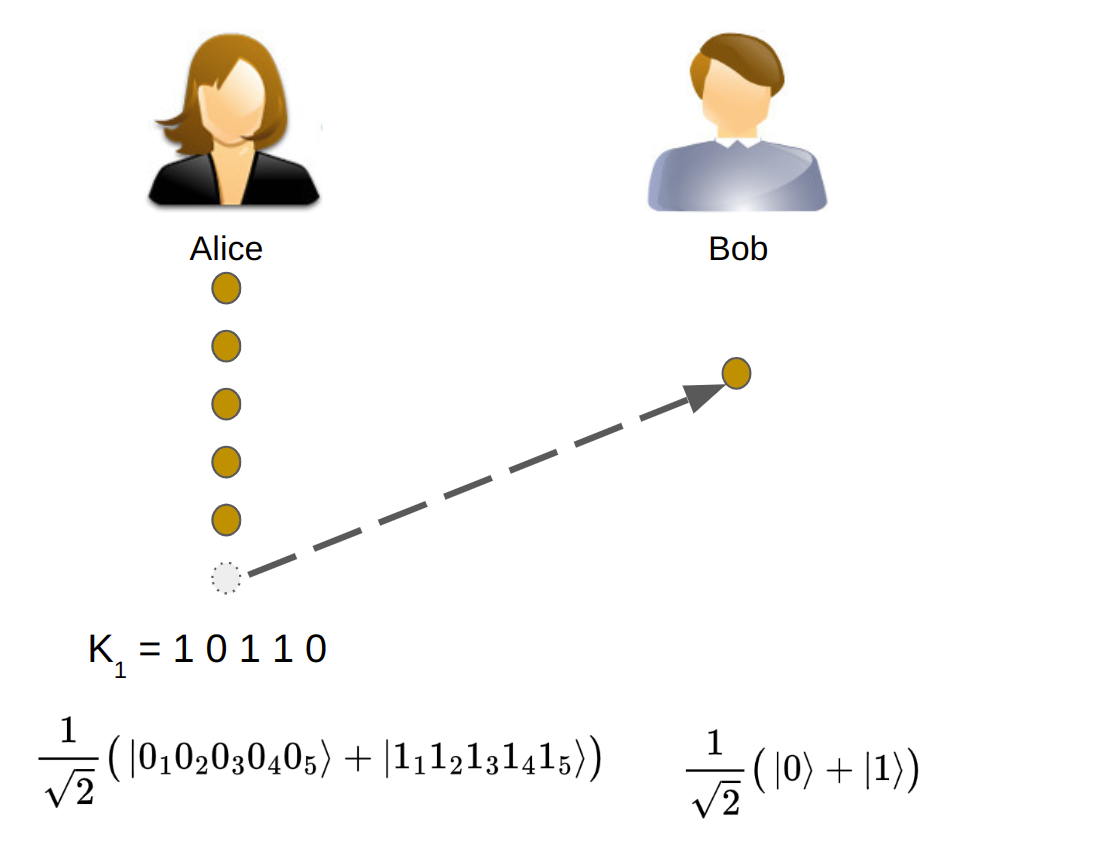}}
\label{fig:step2}}
\hspace{-4mm}
\subfigure[Step 3: Key Transmission]{
\frame{\includegraphics[keepaspectratio=true,angle=0,width=.30\linewidth] {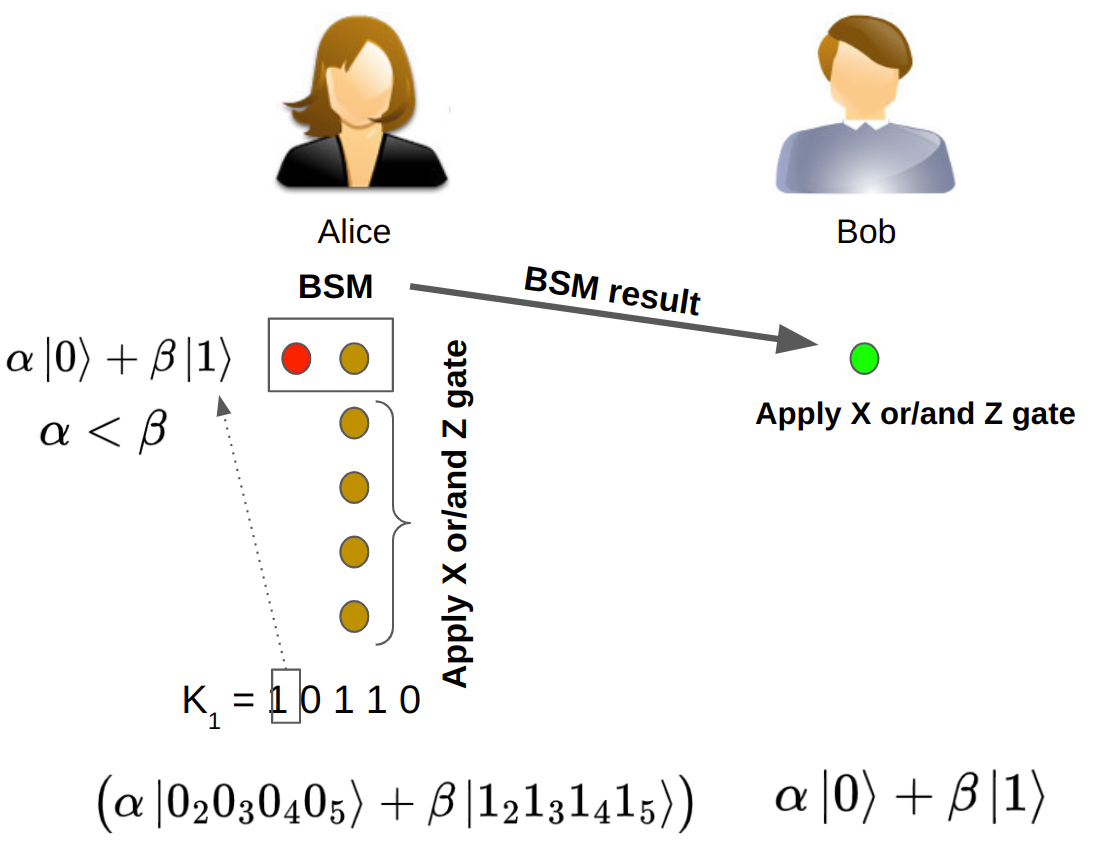}}
\label{fig:step3}}

\subfigure[Step 4: QND Measurement]{
\frame{\includegraphics[keepaspectratio=true,angle=0,width=.30\linewidth] {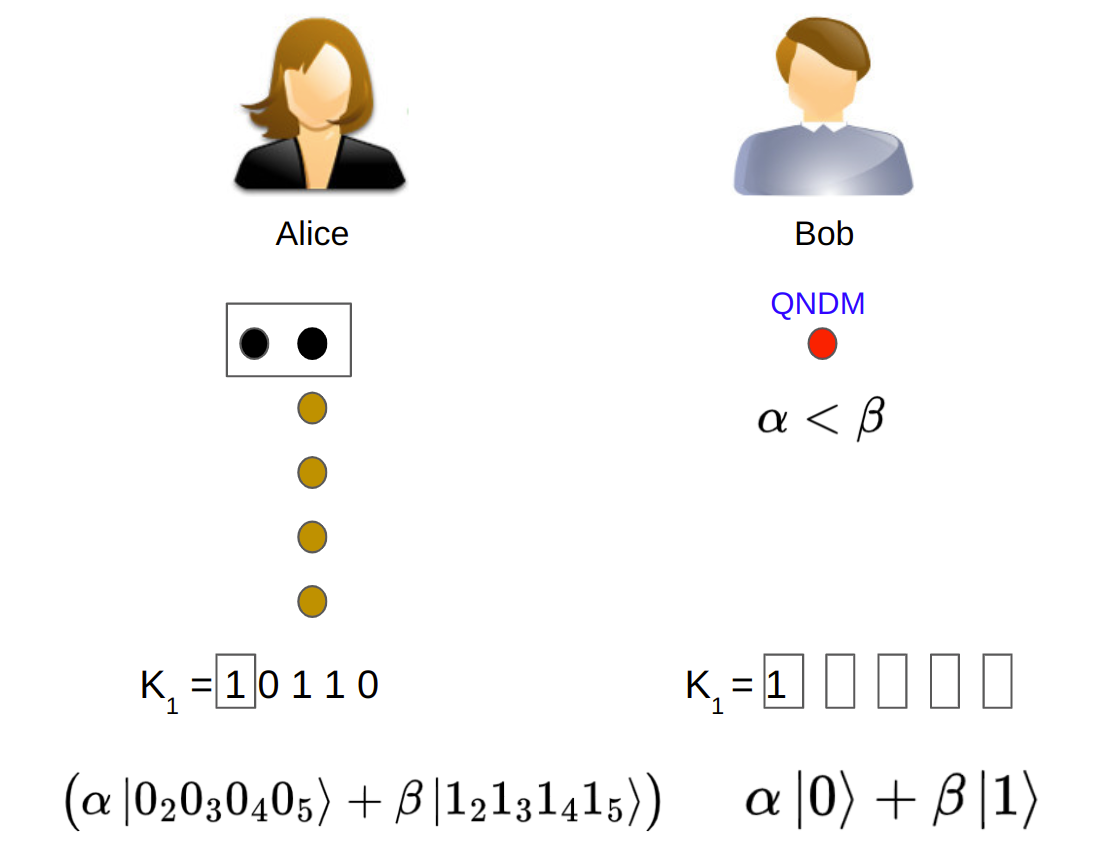}}
\label{fig:step4}}
\hspace{-4mm}
\subfigure[Step 5: GHZ State Reset]{
\frame{\includegraphics[keepaspectratio=true,angle=0,width=.30\linewidth] {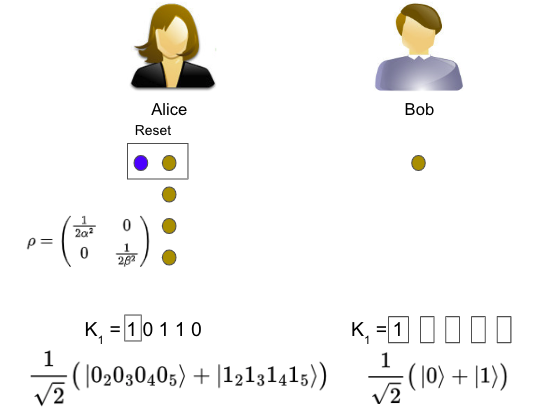}}
\label{fig:step5}}

\caption{Illustration of sending one bit to Bob through our protocol}
\vspace{-6mm}
\label{fig:steps}
\end{center}
\end{figure*}

Alice generates $L + 1$ GHZ state entangled qubits and sends one of them to Bob. She then generates $L$ ancillary qubits and encodes them with the key values that are intended to be transferred to Bob. Alice encodes classical bit $1$ as $\bra{1}$ and $0$ as $\bra{0}$. Alice performs BSM between  the first entangled qubit and the first ancillary encoded qubit and sends the result to Bob. Based on the BSM results, Bob applies desired gates on his qubit and finally measures his qubit using Quantum Nondemolition (QND) measurement. Bob finally reverts the gate operations he performed so that his qubit remains entangled with the $L-1$ qubits that Alice has. Since attackers can intercept the classical communication and learn BSM results, they can estimate the state of the remaining $L-1$ qubits of Alice, thereby making it possible to guess the key. To prevent this, Alice resets the state of the remaining qubits to GHZ state after each BSM measurement with an ancillary bit. Alice and Bob repeat the process until all $L$ bits of Alice are used to transmit the $L$-length key to Bob.



While E91 encodes the ancillary bits with $\ket{0}$ to transmit classical bit $0$ and $\ket{1}$ transmit classical bit $1$, this is not feasible in our solution since the state of the remaining Alice's bits will collapse to $\ket{00\cdots}$ or $\ket{11\cdots}$ after the first BSM. Although they are still entangled, an eavesdropper can see the classical bits and infer the state of the remaining bits. Hence, Alice encodes the classical bit in $\alpha_1\ket{0} + \beta_1\ket{1}$ where $0<\alpha<1$ and $0<\beta<1$. The relationship of $\alpha$ and $\beta$ is used to distinguish between classical bits of $0$ and $1$. In other words, $\alpha > \beta$ is used to transmit classical bit $0$,  and $\alpha < \beta$ is used to transmit classical bit $1$. We next describe the steps in more detail.



\subsection*{Step 1: $L+1$ Qubit GHZ State Preparation} \label{stage:prep}
\begin{figure}
\begin{center}
\includegraphics[keepaspectratio=true,angle=0,width=0.8\linewidth] {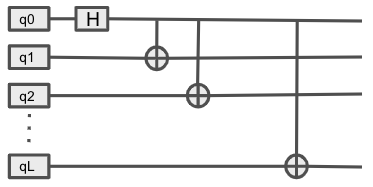}
\vspace{-3mm}
\caption{Preparation of GHZ state with $L+1$ qubits}
\label{fig:GHZ_generation}
\vspace{-6mm}
\end{center}
\end{figure}

Alice prepares $L +1$ qubits in GHZ state. It starts with creating $L+1$ qubits in $\ket{0}$ state. Then, Hadamard and CNOT gates are applied to the qubits to create a GHZ state, as illustrated in Figure~\ref{fig:GHZ_generation}.  The final GHZ state can be written as:
\begin{equation}
\ket{\psi} = \frac{1}{\sqrt{2}}\big(\ket{0_{1}0_{2}0_{3}\cdots0_{L}0_{L+1}} + \ket{1_{1}1_{2}1_{3}\cdots1_{L}1_{L+1}}\big) \label{eqn:standard_GHZ_state}
\end{equation}

In Figure \ref{fig:steps} we considered $L=5$ and in Figure \ref{fig:step1} Alice prepares $5+1$ GHZ state entangled qubits. 

\subsection*{Step 2: Qubit Transmission to Bob} \label{stage:distribute}
Alice keeps the first $L$ qubits (in a quantum memory) and sends the last one to Bob. This requires Alice to have quantum memory with at least $L+1$ qubits capacity; $L$ for GHZ state qubits and $1$ for the ancillary qubit she will use to encode the key. Qubit transmission to Bob can be a direct transmission of the qubit from Alice to Bob if the distance between is short (typically in the order of $130$km or around $80$ miles). Otherwise, quantum repeaters can be used to teleport a qubit with the help of entanglement swapping \cite{pan1998experimentalentangleswapping}. \cite{ji2022entanglementswappbell} describes how qubit teleportation can be achieved using GHZ state qubits. If qubit transmission to Bob fails, Alice will regenerate another qubit, entangle it with her $L$ qubits, and then send it to Bob. In Figure \ref{fig:step2}, Alice sends $6th$ qubit to Bob through a quantum channel.

\subsection*{Step 3: Key Transmission} \label{stage:teleport}
Alice first generates a $L$ bit partial key. She then creates one ancillary qubit, $\theta_1= \alpha_1\ket{0} + \beta_1\ket{1}$, where $\alpha_1 > \beta_1$ if the next bit in the key is $0$ and $\alpha_1 < \beta_1$ if it is $1$. She then performs a Bell State Measurement between the ancillary qubit and the first qubit of the $L$-qubit she has. This will transform the GHZ state to

\begin{multline}
        \ket{0_{\theta1}0_{1}}\big(\alpha_1\ket{0_{2}0_{3}\cdots0_L0_{L+1}} + \beta_1\ket{1_{2}1_{3}\cdots1_L1_{L+1}}\big)\\
            + \ket{0_{\theta1}1_{1}}\big(\beta_1\ket{0_{2}0_{3}\cdots0_L0_{L+1}} + \alpha_1\ket{1_{2}1_{3}\cdots1_L1_{L+1}}\big)\\
            + \ket{1_{\theta1}0_{1}}\big(\alpha_1\ket{0_{2}0_{3}\cdots0_L0_{L+1}} - \beta_1\ket{1_{2}1_{3}\cdots1_L1_{L+1}}\big)\\
            + \ket{1_{\theta1}1_{1}}\big(\alpha_1\ket{1_{2}1_{3}\cdots1_L1_{L+1}} -\beta_1\ket{0_{2}0_{3}\cdots0_L0_{L+1}}\big)\\
\end{multline}

Please note that qubits $0-L$ are located in Alice, whereas qubit $L+1$ is located in Bob. After conducting the BSM using the first qubit of GHZ state and ancillary bit $\theta_1$, she sends the result to Bob using classical communication. Upon receiving the output of BSM, Bob selects which of $X$ and $Z$ gates to apply to the qubit he has. In Figure \ref{fig:step3} Alice prepares an ancillary qubit for the first bit of $k_1$, performs BSM and sends the result to Bob through a classical channel. Bob Applies $X$ or/and $Z$ gate. Alice also applies $X$ or/and $Z$ gate to her qubit to get the same state for every iteration.

\subsection*{Step 4: QND Measurement} \label{stage:measure}
 After applying the appropriate gates, Bob performs QND measurement on his qubit and estimates the value of $\alpha_1$ and $\beta_1$. Since Alice chose the value of $\alpha_1$ and $\beta_1$ based on the value of the classical bit in the key, Bob can infer the bit value based on the estimated $\alpha_1$ and $\beta_1$ value. In other words, Bob infers that the classical bit is $0$ if $\alpha_1>\beta_1$ and $1$ otherwise~\cite{grangier1998quantumQNDoptics}. Unlike the direct measurement of a qubit on the receiver side as done in E91, QND does not disturb the qubit he has, allowing him to reuse it in the following states. However, Bob requires quantum memory that can store its qubit long enough to transmit all $L$ bits in the key. In addition, QND measurement is an approximation of the $\alpha_1$ and $\beta_1$; thus $\alpha_1,\beta_1$ values must be chosen carefully and QND may need to be repeated several times to minimize the error.  Hence, Bob performs QND measurement to his qubit and finds that $\alpha<\beta$ as shown in Figure \ref{fig:step4}. Based on this information, he can extract the first classical bit of $k_1$.

\subsection*{Step 5: GHZ State Reset} \label{stage:reset}
Based on BSM results, Alice applies the same $X$ or $Z$ gates as Bob did to the remaining GHZ state qubits. Alice does this so that the remaining qubits remain in the same state after every teleportation. As a result, the remaining $L$ qubits ($L-1$ in Alice and one in Bob) can be represented as

\begin{equation}
    \ket{\psi}=\alpha_1\ket{0_{2}0_{3}\cdots0_L0_{L+1}} + \beta_1\ket{1_{2}1_{3}\cdots1_L1_{L+1}} \label{eqn:alpha00_beta11}
\end{equation}

To send the second classical bit in the key, Alice encodes another ancillary qubit $\ket{\theta_2} = \alpha_2\ket{0} + \beta_2\ket{1}$ and teleports it to Bob. After the teleportation, the state of qubits will be

\begin{multline}
                \ket{0_{\theta2}0_{2}}\big(\alpha_1\alpha_2\ket{0_{3}0_{4}\cdots0_L0_{L+1}} + \beta_1\beta_2\ket{1_{3}1_{4}\cdots1_L1_{L+1}}\big)\\
            + \ket{0_{\theta2}1_{2}}\big(\alpha_1\beta_2\ket{0_{3}0_{4}\cdots0_L0_{L+1}} + \beta_1\alpha_2\ket{1_{3}1_{4}\cdots1_L1_{L+1}}\big)\\
            + \ket{1_{\theta2}0_{2}}\big(\alpha_1\alpha_2\ket{0_{3}0_{4}\cdots0_L0_{L+1}} - \beta_1\beta_2\ket{1_{3}1_{4}\cdots1_L1_{L+1}}\big)\\
            + \ket{1_{\theta2}1_{2}}\big(\beta_1\alpha_1\ket{1_{3}1_{4}\cdots1_L1_{L+1}} - \alpha_1\alpha_2\ket{0_{3}0_{4}\cdots0_L0_{L+1}}\big)\\
\end{multline}

This is a complex state in the sense that applying only $X$ and $Z$ gate Bob can't extract the value of $\alpha_2$ and $\beta_2$. It would be possible to do so if the state could be reset to the standard GHZ state as in Equation \ref{eqn:standard_GHZ_state} where the probability of being $0$ and $1$ for all qubits is equal. So, we need a reset stage after every teleportation. In Figure \ref{fig:step5} Alice resets her qubit and prepare the remaining qubits in standard GHZ state to send the next classical bit from $k_1$.

We adopt the reset mechanism proposed in \cite{parakh2022quantum1classical}. The proposed solution uses an ancillary qubit and applies a CNOT gate to the target qubit. To reset n-qubit state $\ket{\xi}$ that is currently in the form of $\alpha\ket{00\cdots0} + \beta\ket{11\cdots1}$, in \cite{parakh2022quantum1classical} an ancillary qubit  $\ket{a} = \beta\ket{0} + \alpha\ket{1}$ is used to apply CNOT gate to the first qubit of the $\ket{\xi}$. Next, the first qubit of  $\ket{\xi}$ is measured. If the output is $0$, then the state of  $\ket{\xi}$ is reset to $\frac{1}{\sqrt{2}}\big(\ket{00\cdots0} + \ket{11\cdots1\big)} $ where the ancillary qubit takes the place of the measured qubit. On the other hand, if the measurement output is $1$, $X$ gate is applied to the ancillary qubit before repeating the entire process with a new ancillary qubit. The process is repeated until the measured qubit returns $0$. However, the probability of obtaining $1$ after the first try increases exponentially since $\ket{\xi}$ changes to $\alpha^2\ket{00\cdots0} + \beta^2\ket{11\cdots1}$. So, if $\beta>\alpha$, the coefficient of  $\ket{00\cdots0}$ and $\ket{11\cdots1}$ states will converge to $0$ and $1$, making it impossible to measure $0$. Hence, we modify the solution proposed in \cite{parakh2022quantum1classical} slightly and  encode the ancillary bit $\ket{b}=p\ket{0} + q\ket{1}$, in as a mixed state. Applying the CNOT gate using $\ket{b}$ as the control bit transforms the $\ket{\xi}$ to
\begin{multline}
    \big(p\ket{0}+q\ket{1}\big)\big(\alpha\ket{0_10_2\cdots0_n}+\beta\ket{1_11_2\cdots1_n}\big)\\
    =  p\alpha\ket{00_10_2\cdots0_n} + p\beta\ket{01_11_2\cdots1_n}\\
    + q\alpha\ket{10_10_2\cdots0_n} + q\beta\ket{11_11_2\cdots1_n}
\end{multline}

Now can rewrite the state after measuring the first qubit of $\ket{\xi}$
\begin{multline}
    \ket{0_1}\big(p\alpha\ket{00_2\cdots0_n} + q\beta\ket{11_2\cdots1_n}\big)\\
    \ket{1_1}\big(p\beta\ket{01_2\cdots1_n} + q\beta\ket{10_2\cdots0_n}\big)
\end{multline}

While \cite{parakh2022quantum1classical} uses $p=\beta$ and $q=\alpha$, we prepare this ancillary qubit in a mixed state which can be expressed as the density matrix as  $\rho$ to increase the probability of attaining $0$ from the measurement of the first qubit.

\begin{equation}
        \rho = \begin{pmatrix}
                \frac{1}{2\alpha^2} & 0\\
                0 & \frac{1}{2\beta^2}
                \end{pmatrix} \label{eqn_dm}
\end{equation}

Our experiments show that while the possibility of measuring $0$ is low and sometimes virtually becomes impossible after $3-4$ resets when setting $p$ and $q$ values as described in \cite{parakh2022quantum1classical}, our method guarantees to obtain $0$ (therefore successfully completing the GHZ state reset) in the first try. After the reset, the remaining qubits in Alice are transformed to the standard GHZ state, making it possible to generate another ancillary qubit and apply Steps 1-4 using the next qubit of $\ket{\psi}$ to transmit the next bit in the key until all $L$ bits are transmitted.

\section{Security Analysis}
In this section, we analyze the security of our protocol. We will discuss some common attacks on QKD protocols and show that our protocol is resilient against them. We assume there is an eavesdropper (Eve) who can intercept the classical communications.

\subsection{Entanglement Measure Attack}
Eve has the capability to intercept the transmitted qubit and attempt to entangle an additional qubit with the $(L + 1)$ GHZ state entangled system. Subsequently, Eve can also intercept the Bell State Measurement (BSM) result and carry out step \ref{stage:measure} from her own standpoint. By employing this entanglement attack, Eve aims to acquire the classical bit information, similar to Bob. Nevertheless, it is important to note that this entanglement operation takes place within an expanded Hilbert space \cite{stinespring1955positive}. To detect such eavesdropping attempts, we can examine the violation of the CHSH inequality in the context of $(L+1)$ GHZ state entangled qubits \cite{fan2021greenbergerCHSH}. The concept of \textit{entanglement monogamy}, elucidated by the CKW inequality \cite{coffman2000distributed}, safeguards against a third quantum system (Eve's qubit) becoming maximally entangled with either the $0$-$L$ qubit system on Alice's side or the $(L+1)$th qubit on Bob's side when these two systems are in a maximally entangled state. Consequently, after Eve's entanglement manipulation, Bob's qubit will no longer remain maximally entangled with Alice's qubits. By assessing the violation of the CHSH inequality, Alice and Bob can ascertain whether their system has been subjected to an Eve-mediated attack. 
\subsection{Intercept and Resend Attack}
Eve has the ability to employ an alternative attack technique commonly referred to as the \textit{man-in-the-middle} attack. In this particular context, she intercepts Bob's qubit, performs a measurement on it, and subsequently relays it back to Bob. However, the act of measuring the qubit by Eve introduces perturbations, leading to the collapse of the quantum state into either the $\ket{0}$ or $\ket{1}$ basis states. Consequently, when Bob receives the qubit and carries out the Bell State Measurement (BSM) as an integral step of Alice's teleportation procedure, he fails to obtain the encoded values of $\alpha$ or $\beta$. The conspicuous absence of expected outcomes serves as an immediate indicator for Bob to detect the presence of eavesdropping activity. It is worth noting, however, that our protocol's design effectively mitigates the success of this attack, as step \ref{stage:teleport} is initiated subsequent to the transmission of Bob's qubit. Consequently, Eve's ability to extract meaningful information through this avenue is rendered futile. Moreover, the distinctive features associated with this attack render it readily discernible by Bob.

\subsection{Intercept, Entangle and Resend Attack}
Eve possesses the capability to intercept and withhold Bob's qubit for her own purposes. In order to extend the scope of her eavesdropping activities, she can assemble an additional quantum system consisting of $L+1$ qubits, allocating the $L+1$th qubit to Bob. Consequently, Eve can strategically intercept and analyze each outcome stemming from Alice's Bell State Measurement (BSM). By subjecting her acquired qubit to measurement, Eve can effectively determine the respective values of $\alpha$ and $\beta$, which can subsequently be teleported to Bob. The primary objective underlying this attack pertains to the covert concealment of eavesdropping activities while concurrently extracting comprehensive information. However, it is crucial to note that this attack strategy can be rendered vulnerable to detection through the evaluation of the CHSH inequality violation, akin to the detection methodologies employed in an \textit{Entanglement Measure Attack}.

\section{Performance Analysis}
We verified the correctness of the proposed scheme using the quantum network simulator Netsquid~\cite{coopmans2021netsquid} on a machine with $128$ core AMD EPYC 2.6 GHz CPU and $1$ TiB main memory. Due to memory limitations, we created a GHZ state with up to $13$ qubits and successfully transmitted the $12$ bit key from Alice to Bob. We also demonstrated that it is possible to transfer longer keys by creating multiple GHZ states. For example, if creating a GHZ state with more than $10$ qubit is difficult, one can simply create multiple $10$-qubit GHZ states to transfer as long a key as desired. The source code can be accessed here \textit{https://github.com/HasanTasdiq/GHZ-QKD-Simulation}.

To measure the efficiency of the proposed QKD method, we adopt the formula $\eta = \frac{b_s}{q_t}$ introduced in Cabello~\cite{cabello2000quantumholevo}, where $q_t$ denotes the total quantity of qubits transferred through the quantum channel and $b_s$ represents the total number of classical bits in the secret key. Table \ref{tab:efficiency_comparison} compares the efficiency of different QKD schemes. In our scheme, Alice transmits only one qubit using the quantum channel. Hence,  the efficiency for our scheme is $\frac{L}{1}=L$. In comparison, the efficiency of BB84, E91, and B92 algorithms is less than one as they transmit more qubits than the number of bits in the key. Consequently, the efficiency of our scheme can scale as the number of qubits in a GHZ state increases. As previous studies described how to create a GHZ state with an up to $2,000$ qubits~\cite{zhao2021GHZcreation2000,mooney2021generation27,mooney2021whole65}, the efficiency score of the proposed algorithm can be as high as $2,000$.

\begin{table}
\centering
\caption{Efficiency, $\eta$, of different QKD schemes. $q_t$ refers to the total quantity of qubits transferred through the quantum channel and $b_s$ represents the total number of classical bits in the secret key.}
\vspace{-2mm}
\label{tab:efficiency_comparison}
\begingroup
\setlength{\tabcolsep}{2pt} 
\begin{tabular}{c c c c}
\hline

Scheme  &  $b_s$ & $q_t$ & $\eta$\\
\hline
Bennett, 1992 \cite{bennett1992quantumB92} & $<0.5$ & 1 & $<0.5$ \\
\hline
Bennett and Brassard, 1984 \cite{bennett2020quantumbb84} & 0.5 & 1 & 0.5\\
\hline
Goldenberg and Vaidman, 1995 \cite{goldenberg1995quantum} & 1 & 2 & 0.5 \\
\hline
Ekert, 1991 \cite{ekert1991quantumE91} & 1 & 1 & 1\\
\hline
Koashi and Imoto, 1997 \cite{koashi1997quantum} & 1 & 2 & 0.5 \\
\hline
Cabello, 2000  \cite{cabello2000quantum} & 2 & 2 & 0.5 \\
\hline
Our scheme & L & 1 & $L, (L>1)$ \\
\hline
\end{tabular}
\endgroup
\vspace{-3mm}
\end{table}
\section{Extensions Of The Model}
\subsection{Multi Party Quantum Key Distribution} \label{extension: MPQKD}
To broadcast a secret key for a multi-party QKD, Alice sends the key to $n$ parties, $Bob_1, Bob_2\cdots Bob_n$. To do so, we modify Step 1 (i.e., Preparation Stage)~\ref{stage:prep} for Alice to prepare $L+n$ qubits and send $n$ qubits to the $n$ parties. Alice performs BSM with the ancillary qubit using her $ith$ entangled qubit and sends the result to all the $n$ parties. All $n$ parties perform QND as described in the Measurement Stage~\ref{stage:measure} to extract the classical bit information at the same time.

\subsection{High Capacity Server and Low Capacity Client}
In our scheme, Alice needs to have high quantum resources to keep $L$ entangled qubits and to generate $2L$ ancillary qubits. To enable Quantum key sharing between two low-capacity nodes, we introduce another version of our scheme as High Capacity Server and Low Capacity Client where clients ask for a secret key from a centrally-managed high-capacity server. The proposed multi-party quantum key distribution scheme (as discussed in Section~\ref{extension: MPQKD}) can be used to implement the high-capacity GHZ state generator. Specifically, Alice can act as the high-capacity GHZ state generator to produce a multi-qubit GHZ state and allow the end users to access a common key generated by the server.

\section{conclusion and future work}
Quantum Key Distribution (QKD) is one of the important use cases of quantum networks as it can provide a means to transmit encryption keys in an attack-proof manner. The state-of-the-art QKD algorithms, such as BB84 and E91, requires too many qubit transmission between end users. This, in turn, can be a limiting factor due to the limited capacity of quantum networks. In this work, we present a multi-qubit Greenberger–Horne–Zeilinger State-based QKD scheme to lower the need for qubit transmissions in QKD. The proposed novel solution requires as little as one qubit transmission between the users to transmit encryption keys. Through simulations, we demonstrate that the proposed scheme is a viable alternative to existing QKD solutions. We further discuss possible attack scenarios and present a defense mechanism based on readily available methods such as CHSH inequality. Since the proposed method requires the transfer of more information through classical network channels,  we will explore ways to minimize the number of shared classical bits in future work. We will further investigate possible methods to minimize the number of gate operations applied to minimize the error probability due to noise in gate operations.


\bibliographystyle{IEEEtran}
\footnotesize

\bibliography{references}
\end{document}